\documentstyle [11pt]{article}
\def\fnote#1#2{\begingroup\def\thefootnote{#1}\footnote{#2}\addtocounter
{footnote}{-1}\endgroup}

\begin{document}

\hfill{UTTG-06-13}

\vspace{36pt}

\begin{center}
{\large {\bf {Tetraquark Mesons in Large-$N$ Quantum Chromodynamics}}}

\vspace{36pt}
Steven Weinberg\fnote{*}{Electronic address:
weinberg@physics.utexas.edu}\\
{\em Theory Group, Department of Physics, University of
Texas\\
Austin, TX, 78712}

\vspace{30pt}

\noindent
{\bf Abstract}
\end{center}

\noindent
It is argued that exotic mesons consisting of two quarks and two antiquarks are not ruled out in quantum chromodynamics with a large number $N$ of colors, as generally thought.  They can come in two varieties:  short-lived tetraquarks with decay rates proportional to $N$, which would be unobservable if $N$ were sufficiently large, and long-lived tetraquarks with decay rates proportional to $1/N$.  The $f_0(500)$ and $f_0(980)$ may be examples of these two varieties of exotic mesons.

\vfill

\pagebreak

The suggestion[1]  to consider quantum chromodynamics in the limit of a large number $N$ of colors, with the gauge coupling $g$ vanishing in this limit as $1/\sqrt{N}$,  has been had some impressive success in reproducing qualitative features of strong interaction phenomena.  In his classic Erice lectures[2] describing these results, Coleman concluded that, for large $N$, quantum chromodynamics should not admit tetraquark mesons --- exotic mesons that are formed from a pair of quarks and a pair of antiquarks --- a result that has been widely accepted[3].  This note will complete Coleman's argument and then point out exceptions to his conclusion.  The large $N$ approximation not only does not rule out tetraquark mesons, it helps to understand their properties.

Coleman's reasoning was as follows.  By Fierz rearrangements of fermion fields, any color-neutral operator formed from two quark and two antiquark fields can be put in the form 
\begin{equation}
{\cal Q}(x)=\sum_{ij}C_{ij}{\cal B}_i(x){\cal B}_j(x)\;,
\end{equation}
where the ${\cal B}_i(x)$ are  various color-neutral quark bilinears:
\begin{equation}
{\cal B}_i(x)=\sum_a\overline{q^a (x)}\Gamma_i q^a(x)\;.
\end{equation}
Here $q^a$ is a column of canonically normalized quark fields,\fnote{**}{In his article Coleman used bilinears  ${\cal B}'_i(x)$  defined to contain an extra factor of $g^{2}N^{1/2}\propto N^{-1/2}$.  This makes no difference to results for observables.} with $a$ an $N$-component $SU(N)$ color index and with spin and flavor indices suppressed; the $\Gamma_i$ are various $N$-independent spin and flavor matrices; and the $C_{ij}$ are some 
symmetric numerical coefficients, which we will take as $N$-independent.    Coleman considered the vacuum expectation value of two of these fields, given by a decomposition into disconnected and connected parts
\begin{eqnarray}
&&\left\langle {\cal Q}(x){\cal Q}(y)\right\rangle_0=\sum_{ijkl}C_{ij}C_{kl}\Bigg[\left\langle {\cal B}_i(x){\cal B}_k(y)\right\rangle_0\left\langle {\cal B}_j(x){\cal B}_l(y)\right\rangle_0 \nonumber\\&&~~~+\left\langle {\cal B}_i(x){\cal B}_j(x) {\cal B}_k(y){\cal B}_l(y)\right\rangle_{0,{\rm conn}}\Bigg]\;.~~~~~~~
\end{eqnarray}
(We drop disconnected terms that are coordinate-independent.)
A one-tetraquark pole can only appear in the final, connected, term, but according to the usual rules for counting powers of $N$, the first  term is of order $N^2$, while the final term is only of order $N$, and so any one-tetraquark pole would make a contribution in (3) that is relatively suppressed by a factor $1/N$.

So far, so good, but what does this really show?  Coleman concluded ``In the large-$N$ limit, quadrilinears make meson pairs and nothing else.'' But is this justified?  If   there is a tetraquark meson pole in the connected part of the propagator, what difference does it make if its residue is small compared with the disconnected part?  To take an analogy, the amplitude for ordinary meson-meson scattering is proportional to the connected part of a four-point function involving four quark-antiquark bilinear operators, which is of order $N$, while the disconnected parts of the same four-point function are of order $N^2$.  Does this mean that ordinary mesons do not scatter in the large $N$ limit?

The real question is the decay rate of a supposed tetraquark meson.  If the width of the tetraquark grows as some power of $N$, while its mass is independent of $N$, then for very large $N$ it may not be observable as a distinct particle.  Although Coleman did not address this issue, his discussion does suggest that the rate for an tetraquark meson to decay into two ordinary mesons does grow with $N$.  As we will now see, this is correct, but with an important exception.

To calculate decay rates, we need to represent particle states with operators that are properly normalized to be used as LSZ interpolating fields.  
The propagator for a quark bilinear operator ${\cal B}_n(x)$ representing an ordinary meson is proportional to $N$, but the residue of the pole in the propagator of a properly normalized operator should be $N$-independent, so as noted by Coleman, the properly normalized operators for creating and destroying ordinary mesons are $ N^{-1/2}{\cal B}_n(x)$.  Similarly, if there is an  one-tetraquark pole in the connected term in (3), then since the connected term in Eq.~(3) is of order $N$, the correctly normalized operator for creating or destroying a tetraquark meson is $N^{-1/2}{\cal Q}(x)$.   The amplitude for the decay of a tetraquark  meson into ordinary mesons of type $n$ and $m$ is then proportional to a suitable Fourier transform of the three-point function
\begin{eqnarray}
&&N^{-3/2}\left\langle T\{{\cal Q}(x){\cal B}_n(y){\cal B}_m(z)\}\right\rangle_0\nonumber\\&&=
N^{-3/2}\sum_{ij}C_{ij}\left\langle T\{{\cal B}_i(x){\cal B}_n(y)\}\right\rangle_0 \left\langle T\{{\cal B}_j(x){\cal B}_m(z)\}\right\rangle_0\nonumber\\&&~~~~~~+N^{-3/2}\left\langle T\{{\cal Q}(x){\cal B}_n(y){\cal B}_m(z)\}\right\rangle_{0,{\rm conn}}\;.
\end{eqnarray}
The connected second term on the right is of order $N^{-3/2}N=N^{-1/2}$, but the first term is larger, 
of order $N^{-3/2}N^2=N^{1/2}$, giving a decay rate proportional to $N$.  In this case a tetraquark meson would become unobservable for $N\rightarrow \infty$, though one may still wonder about the relevance of this result.  The physical value $N=3$ may or may not be taken as large, but it can't be regarded as infinite.

In any case, there is an exception to the rule that tetraquark mesons become increasingly unstable for increasing $N$.  It may be that  the bilinears ${\cal B}_i(x)$ in Eq.~(1) have quantum numbers that do not match those of any meson light enough to appear in the decay of the tetraquark meson represented by ${\cal Q}(x)$.  In that case the tetraquark  decay amplitude would arise entirely from the second term in Eq.~(4), which would give a decay rate of the tetraquark into two light ordinary mesons  proportional to $1/N$, just as in the decay of ordinary mesons.  

We can find examples of both kinds.  For an example of a short-lived tetraquark, consider a   $J^{PC}=0^{++}$ isoscalar tetraquark meson, represented by the operator
\begin{equation}{\cal Q}(x)=\sum_{ab}\Big(\overline{q^a (x)}\,\gamma_5 \,\vec{t}\, q^a(x)\Big)\cdot 
\Big(\overline{q^b (x)}\,\gamma_5\, \vec{t}\, q^b(x)\Big)\;,\end{equation}
where $\vec{t}$ is an isospin matrix.  In this case the decay into two pions can proceed through the first term in Eq.~(4), and the decay rate is of order $N$.  This may be the case for a plausible tetraquark[4], the very broad $f_0(500)$, which has a width for two-pion decay of 400 to 700 MeV.

For an example of a long-lived tetraquark,  consider the case of a different $J^{PC}=0^{++}$ isoscalar tetraquark meson represented by the operator
\begin{equation} {\cal Q}(x)=\left|\sum_a\overline{u^a (x)}\gamma_5 u^a(x)+\sum_a\overline{d^a (x)}\gamma_5 d^a(x)\right|^2\;,
\end{equation}
 or 
\begin{equation} {\cal Q}(x)=\left|\sum_a\overline{s^a (x)}\gamma_5 u^a(x)\right|^2+
\left|\sum_a\overline{s^a (x)}\gamma_5 d^a(x)\right|^2\;.\end{equation}
 The lightest meson with the quantum numbers  of these choices of  ${\cal B}(x)$ are the $\eta(548)$ and  the $K(495)$.  If a $0^{++}$ tetraquark meson represented by (6) or (7) is lighter than $2m_\eta$ or $2m_K$, respectively,  its decay would receive no contribution from the first term in Eq.~(4).  Its decay amplitude would arise entirely from the second term in Eq.~(4), which would give a decay rate (for instance, into two pions) proportional to $1/N$.  This may be the case for instance for the $f_0(980)$, which is plausibly identified as a tetraquark meson[4], and has a width of only 40 to 100 MeV.  The large $N$ approximation not only does not rule out such exotic messons --- it can explain why they are narrow.

The large $N$ approximation gives an objective meaning to a statement that a tetraquark represented by a product ${\cal B}_1(x){\cal B}_2(x)$ of quark bilinears is a composite of the ordinary mesons represented by ${\cal B}_1(x)$ and ${\cal B}_2(x)$, even where the tetraquark meson is much lighter or much heavier than the sum of these ordinary meson masses.    It is not only that the two-meson intermediate state dominates the propagator of the tetraquark operator ${\cal Q}(x)$, as shown by Coleman.  More relevant to experiment,  the contribution of a two-meson state to meson--meson scattering is proportional to $[(N^{-1/2})^4N]^2=1/N^2$, while since the amplitude for the tetraquark to go into two ordinary mesons is proportional to $N^{1/2}$, the contribution of a tetraquark pole (if there is one) is proportional to $[N^{1/2}]^2=N$.  Hence, whatever its mass, for large $N$ the one-tetraquark intermediate state dominates the scattering of these two ordinary mesons in the partial wave with the same quantum numbers as the tetraquark.

It would be interesting to apply this analysis to  a wider variety of tetraquarks, with  quantum numbers other than $0^{++}$, $T=0$, and also taking flavor $SU(3)$ symmetry into account.

\vspace{20pt}

I am grateful to  Frank Close and Philip Page for helpful correspondence, and to Tamar Friedmann for a seminar talk that spurred my interest in tetraquarks.  This material is based upon work supported by the National Science Foundation under Grant Number PHY-0969020 and with support from The Robert A. Welch Foundation, Grant No. F-0014.

\begin{center}
{\bf ----------}
\end{center}

\vspace{10pt}

\begin{enumerate}
\item G. 't Hooft, Nucl. Phys. {\bf B75}, 461 (1974).
\item S. Coleman, {\em Aspects of Symmetry} (Cambridge University Press, Cambridge, UK, 1985), pp. 377-378.
\item For instance, see P. R. Page, in {\em Intersections of Particle and Nuclear Physics: 8th Conference}, ed. Z. Parsa (American Institute of Physics, 2003), p. 513.
\item R. J. Jaffe, Phys. Rev. D {\bf 15}, 267 (1977); F. E. Close and N. A. T\"{o}rnqvist, J. Phys. G {\bf 28}, R249 (2002) [hep-ph/0204205]; T. Friedmann,  to appear in Eur. Phys. J. C [arXiv:0910.2229]. 
\end{enumerate}

\end{document}